\documentclass[twocolumn,showpacs,aps,amsmath,amssymb,prl,preprintnumbers]{revtex4}

\newcommand{\BE}{\begin{equation}}
\newcommand{\EE}{\end{equation}}
\newcommand{\BA}{\begin{eqnarray}}
\newcommand{\EA}{\end{eqnarray}}
\usepackage{graphicx}% Include figure files
\usepackage{dcolumn}% Align table columns on decimal point
\usepackage{bm}% bold math
\input epsf.tex

\begin{document}

\title{Band gap or negative refraction effects?}

\author{Chao-Hsien Kuo}\author{Zhen Ye}%\email{zhen@phy.ncu.edu.tw}
\affiliation{Wave Phenomena Laboratory, Department of Physics,
National Central University, Chungli, Taiwan}

\date{October 16, 2003}

\begin{abstract}

By a rigorous numerical simulation based on the standard multiple
scattering theory, we investigate the optical transmission in
photonic crystal structures, formed by dielectric cylinders
embedded in parallel in a uniform medium. In contrast to previous
conjectures, the results indicate that the imaging effect which
has been interpreted as a signature of the negative refractive
effects is caused by a tunnelling effect in the presence of
partial band gaps. There is no need for an all-angle negative
refraction effect to understand some peculiar phenomena of
photonic crystals.

\end{abstract}

 \maketitle

Ever since the proposal that a perfect lens can be realized by the
so called Left Handed Material (LHM) or Negative Refraction Index
Material (NRIM), a conceptual material first introduced by
Veselago many years ago \cite{Ves}, the research on such a super
lens and LHM has been skyrocketing in the mist of much debate. A
great body of literature has been and continues to be generated.

Although there are a few recent challenges with regard to the
concept of LHM or relevant negative refraction effects
\cite{Valanju,Efros,Hooft,comm,Garcia,Ye}, the mainstream
consensus has been that some indications of negative refraction
effects are affirmative. Supports from both theoretical and
experimental perspectives have been reported
\cite{PRB,Nature,Theo,exp,Chuang}. Among them, the major portion
of research has been on generating the negative refraction effects
by photonic crystals (PCS)
\cite{PRB,Nature,exp,Chuang,Kosaka,APL}. For example, the authors
in \cite{PRB} demonstrated an unusual focused image when
electromagnetic waves propagate through a rectangular slab of
photonic crystals formed by regular arrays of dielectric cylinders
in a uniform medium. Since the image could not be explained in the
framework of the usual positive refraction, the authors have
attributed the cause to an all-angle negative refraction effect.

Upon inspection, we found that the claim of previous supports for
the negative refraction effects generated by photonic crystals is
questionable. As a matter of fact, we found that the unusual
imaging or the apparent abnormal refraction that has been thought
to be the negative refraction effects can be well explained in
terms of the partial gaps revealed by the photonic crystals. In
the present Letter, we show some results to support our point of
view.

Here we consider the transmission of photonic waves in photonic
crystals. To be more relevant with published results, we will use
the photonic crystal structures which have been commonly adopted
in previous simulations, such as those in \cite{PRB,Nature}.
Unlike most previous FDTD simulations, we will employ the standard
multiple scattering theory to compute the propagation and
scattering of the waves. This theory is exact and was first
formulated systematically by Twersky \cite{Twersky}, then has been
reformulated and applied to optical, sonic and water wave problems
\cite{JOSA,Chen,Yue}. In guiding our discussion, the band
structures of the photonic crystals will be computed by the
conventional plane wave expansion method. As we will see, an image
can indeed appear across a rectangular slab of photonic crystals.
But this image is not due to the negative refraction effect,
rather it is caused by guided propagation in the presence of a
partial gap in the corresponding band structures.

The systems we consider are two dimensional arrays of parallel
dielectric cylinders placed in a unform medium, which we assume to
be air. Consider an arbitrary array of identical dielectric
cylinders placed in air. The solution for the wave scattering or
propagation through such an array can be obtained by the multiple
scattering theory. The essence of the theory is summarized as
follows. In response to the incident wave from the source and the
scattered waves from other scatterers, each scatterer will scatter
repeatedly waves and the scatterered waves can be expressed in
terms of a modal series of partial waves. Considering these
scattered waves as an incident wave to other scatterers, a set of
coupled equations can be formulated and computed rigorously. The
total wave at any spatial point is the summation of the direct
wave from the source and the scattered waves from all scatterers.
The intensity of the waves is represented by the modulus of the
wave field.

For brevity, we only consider the E-polarized waves, that is, the
electric field is kept parallel to the cylinders. The following
parameters are used in the simulation. (1) The dielectric constant
of the cylinders is 14, and the cylinders are arranged in air to
form a square lattice. (2) The lattice constant is $a$ and the
radius of the cylinders is 0.3$a$; in the computation, all lengths
are scaled by the lattice constant.

First we consider the propagation of photonic waves through a
rectangular slab of an array of dielectric cylinders, by analogy
with those shown in \cite{PRB,Nature}. The slab width (vertical)
equals 40$\sqrt{2}$, and the length (horizontal) is 3$\sqrt{2}$.
The source is placed at 0.35 from the left side of the lattice -
the geometry can be referred to in Fig.~1. The incident wave
propagates along the $\Gamma M$, i.~e. the [1,1] direction.

Figure~1 shows the imaging fields and the band structure. On the
left panel, the band structure is plotted and the qualitative
features are similar to that obtained for a square array of
alumina rods in air \cite{Nature}. The horizontal darken area
refers to the region in which the frequency band is convex. This
frequency range, called AANR region in \cite{PRB}, has been
regarded to be essential for all angle negative refraction
\cite{PRB}. Plotted on the right panel are the real part as well
the modulus of the total electrical field. The frequency is chosen
at 0.192$*2\pi c/a$ which is nearly at the middle of the `AANR'
region. Here, we indeed observe a focused image point on the right
hand side of the slab. This imaging effect is more apparent in the
plot of the real part of the field, and is essentially the same as
that observed by the FDTD simulation for the similar photonic
crystals \cite{PRB}. Moreover, such a focusing effect is
persistent for frequencies within the shaded area and also a
regime just outside this area \cite{Bikash}; we note here that the
data interpretation in Ref.~\cite{Bikash} is not appropriate.

In the conventional thought, such an focused imaging property is
impossible. The reason follows. At the frequency considered, the
lattice constance is much smaller than the wavelength of the
incident wave (i.~e. the ratio amounts to 0.192). Therefore the
transmitted wave might be expected to be insensitive to the
detailed structure of media, that is, the homogenization effect.
Therefore the slab may be regarded as an effective medium with an
effective refractive index. If the refractive index were positive,
the transmitted waves would diverge. As such, no image could be
possibly focused on the other side of the slab of materials. This
scenario is depicted by Fig.~1 (b1). Thus the focused image shown
by Fig.~1(c) cannot be explained in terms of the positive
refraction. In the literature, such a focused imaging effect has
been believed to be the onset of negative refraction, and has been
thought to be the major evidence of LHM (e.~g.
Refs.~\cite{PRB,Nature,Chuang}). Indeed, if the refraction at an
interface is negative, the refracted transmission will be
deflected towards the same side as the incident wave with respect
to the normal of the interface. Such an unusual deflection can
give rise to a focused image when waves pass through a rectangular
slab. This negative refraction scenario is illustrated by
Fig.~1(b2).

Though tempting, there is ambiguity in the explanation of the
apparent focusing effect in the context of negative refraction. As
indicated in Fig.~1(b2), if the focused image is caused by the
effectively negative refraction, particularly when the negative
refraction is all angle, a focused point should also prevail
inside the slab, according to Fig.~1 (b2). The expected focused
image, however, is not noticeable neither in the previous
simulations, nor in our simulations.

When we further explore the focusing effect shown on the other
side of the slab, we found that such focusing effect does not have
to be explained as the occurrence of negative refraction. Rather,
it is purely due to the nature of some band structures. The band
structure from Fig.~1(a) clearly shows that in the frequency range
considered, there is a band gap along the $\Gamma X$ direction,
i.~e. the [10] direction. Therefore the waves are prohibited from
propagation along this direction. In other directions such as
$\Gamma M$, however, there is an allowed band to support the
propagation of waves. In the present setup, as well as in previous
setups, the incident wave is set along the $\Gamma M$ direction,
which makes an angle of 45 degrees to the $\Gamma X$ direction. As
prohibited from propagating in the $\Gamma X$ direction, i.~e. 45
degrees from the straight horizontal direction, waves naturally
tend to move forward along the $\Gamma M$ direction. The frequency
band in the $\Gamma M$ direction provides a propagating avenue for
the waves to go over to the other side of the slab, like a gas
pipeline or a water tap. To put our discussion into perspective,
we have done a variety of simulations. Some of the key results are
shown below.

Fig.~2 shows the imaging fields with a larger rectangular slab.
All the parameters are taken as the same as that in Fig.~1, except
that the horizontal length of slab has been increased to
$20\sqrt{2}$, allowing us to study the nature of wave propagation
inside the slab. The overall fields are imaged in (a1) and (a2) in
terms of the real part and the modulus of the electric field
respectively. The fields within the slab have been purposely
zoomed and replotted in (b1) and (b2) for the sake of clarity.
Fig.~2(a1) clearly shows a focused image across the slab. Within
the slab, it is clearly shown by (a2), (b1) and (b2) that the
transmission is mainly focused within a pipeline or tunnel along
the $\Gamma M$ direction. When plotting in the real time, such a
guided transmission is more beautifully presented. This feature is
fully in accordance with the above discussion of the expected
properties of the band structure that has a partial gap. The
apparent focused image across the slab is in fact just the
outburst point of the transmitted waves. To put it simple, the
passing band in the $\Gamma M$ acts as a transportation carrier
that moves the source to the other side of the slab. The waves on
the right hand side of the slab looks as if they were radiated by
an image that has been transported across the slab within a narrow
guide. If such an imaging phenomenon had to be interpreted in the
framework of an effective medium theory without worrying about
what is really going on inside the medium, the negative refraction
would be one of the options to be resorted to. But the present
results suggest that this be an artifact explanation. In fact, we
found that many published works, such as Ref.~\cite{Chuang}, have
not considered what has been going on inside the transporting
media. In fact, the experimental evidence of the negative
refraction in Ref.~\cite{Chuang} can be explained without the need
for the refraction to be negative. We will show our discussion
elsewhere.

To further support our observation, we place a transmitting source
{\it inside} an array of cylinders. The overall shape of the array
is square. Fig.~3 presents the simulation results. Again, all the
parameters for the physical quantities are taken from Fig.~1. The
cylinders are arranged to form a square crystal inside a square
area whose side measures as 14$\sqrt{2}$. The geometry of the
setup can be seen in Fig.~3. Here, to show the results in their
most explicit form, we plot separately the fields within and
outside the photonic crystal structure. Fig.~3 (a1) and (a2) show
the real part and the modulus of the electric field outside the
array of the cylinders, whereas Fig.~3 (b1) and (b2) show the real
part and the modulus of the electric field inside the array of the
cylinders. As expected, the focused images are evident in four
allowed directions, in the [11], [-1,1], [-1,-1], and [1,-1]
directions respectively, as shown by (a1) and (a2). The images of
the field inside the array clearly show the travelling path of the
waves along these directions, depicted by (b1) and (b2).

We have also considered the situation in which we place the
transmitting source inside an array of cylinders and the array
takes roughly the circular shape. The results are shown in Fig.~4.
Again, if the effectively-negative refraction or all-angle
negative refraction exists, the waves are expected to propagate in
all directions - no focusing should be expected. The results
indicate that this is untrue. Similar to plotting Fig.~3, the
geometry of the setup can be seen in Fig.~4. The fields within and
outside the photonic crystal structure are plotted separately.
Fig.~4 (a1) and (a2) show the real part and the modulus of the
electric field outside the array of the cylinders, while Fig.~4
(b1) and (b2) show the real part and the modulus of the electric
field inside the array of the cylinders. Here we also observe the
focused images as in the case of Fig.~3. These images are in the
[11], [-1,1], [-1,-1], and [1,-1] directions respectively, as
shown by (a1) and (a2). Inside the arrays, the waves clearly
travel along these directions, depicted by (b1) and (b2). By
comparing Fig.~3 and Fig.~4, we conclude that the imaging feature
and wave propagation are irrelevant to the outer shape of arrays,
i.~e. the boundary of the photonic crystals.

All the imaging effects shown so far can be well explained in
terms of the band structure properties. We have also considered
other frequencies both inside and outside the shaded area of
Fig.~1 (a). We observe that the field patterns are qualitatively
similar. But when the frequency moves towards to the edge of the
partial band gap in the $\Gamma X$ direction, the apparent guided
propagation inside the photonic crystal tends to diminish, leading
to a gradual disappearance of the `focusing' effects. This is
reasonable. As the frequency moves towards the edge of the
forbidden band of the $\Gamma X$ direction, the waves would be
either allowed to propagation in all direction or forbidden to
travel in all direction due to the complete band gap located just
above the first frequency band. Therefore the imaging effect
caused by the partial band gap diminishes.

In summary, we have considered the optical transmission in a
photonic crystal. The results show that the imaging effect which
has been previously interpreted as a signature of the negative
refractive effects is caused by a tunnelling or guided propagation
in the presence of partial band gaps. The negatively interpreted
refraction needs not to be negative.

{\bf Acknowledgements} The work received support from NSC and NCU.
The early participation of Mr. Ken Kang-Hsien Wang is greatly
appreciated. One of us (CK) is particularly grateful to Mr. Wang
for many inspiring discussions and help.

\newpage

\section*{Figure Captions}

\begin{description}

\item[Figure 1] \label{fig1} Left panel: (a) Band structure; the
shaded horizontal region indicates the frequency range within
which the frequency band is everywhere convex; (b) the conceptual
diagrams showing the conventional and negative refraction
scenarios. Right panel shows the imaging fields across the slab:
(c) The real part of the electric field; (d) The intensity field,
i.~e. the modulus of the electric field in the log scale.
Hereafter, the dark circles denote the cylinders. The source is on
the left hand side of the slab.

\item[Figure 2]  \label{fig2} The imaging fields for a slab of
photonic crystal structure, similar to Fig.~1 but with a larger
sample length. (a1) and (b1) show the real part of the field, (a2)
and (b2) show the intensity field. (b1) and (b2) are replots of
the imaging fields for the area inside the slab from (a1) and (a2)
respectively. Here we see clearly that the waves propagate in a
small tunnel inside the slab; it is most evident in (b2).

\item[Figure 3]  \label{fig3} The imaging fields for a
transmitting source located inside a square array of cylinders.
The square measures as $14\sqrt{2}\times 14\sqrt{2}$. All other
parameters are the same as in Fig.~1.

\item[Figure 4]  \label{fig4} The imaging fields for a
transmitting source located inside a roughly circularly shaped
array of cylinders. The total number of cylinders is 484. All
other parameters are the same as in Fig.~1.

\end{description}

\end{document}